\documentclass[conference]{IEEEtran}
\usepackage{cite}
\usepackage{amsmath,amssymb,amsfonts}
\usepackage{algpseudocode}
\usepackage{graphicx}
\usepackage{textcomp}
\usepackage{xspace}
\usepackage{xcolor}
\usepackage[numbers]{natbib}
\usepackage{amsmath}
\usepackage[bookmarks=false]{hyperref}
\usepackage[nameinlink, capitalise]{cleveref}
\usepackage{pgfplots}
\usepackage[section]{placeins}
\usepackage{multirow}
\usepackage{etoolbox}
\usepackage{caption}
\usepackage{ifthen}
\usepackage{hhline}
\usepackage{xspace}
\usepackage{subcaption}
\usepackage{cite}
\usepackage{booktabs}
\usepackage{adjustbox}
\usepackage{paralist}
\usepackage{flushend}

\usepackage{tikz}
\usepackage{pgf-umlcd}
\usetikzlibrary{arrows.meta,shapes,positioning, calc, fit, shadows,decorations.pathreplacing}
\usetikzlibrary{shapes,calc,fit}
\usetikzlibrary{automata,positioning}
\usetikzlibrary{calc, chains, intersections, shapes, trees}
\usepackage{bm}
\usepackage{algorithm}
\usepackage{varwidth}
\usepackage{siunitx} 
\usepackage{todonotes}
\usepackage{pdfpages}
\usepackage{siunitx}
\usepackage{acro}
\usepackage{fancyhdr}

\crefname{figure}{Fig.}{Figs.}

\usepackage[font=small, justification=justified,singlelinecheck=true,tableposition=top]{caption}

\makeatletter
\def\endthebibliography{%
	\def\@noitemerr{\@latex@warning{Empty `thebibliography' environment}}%
	\endlist
}

\acsetup{
    single=true,
    case-sensitive=false
}

\DeclareAcronym{arp}{
  short = ARP,
  long  = address resolution protocol
}

\DeclareAcronym{dds}{
  short = DDS,
  long  = Data Distribution Service
}

\DeclareAcronym{dma}{
  short = DMA,
  long  = Direct Memory Access
}

\DeclareAcronym{rdma}{
	short = RDMA,
	long  = Remote Direct Memory Access
}

\DeclareAcronym{ebpf}{
  short = eBPF,
  long  = extended Berkeley packet filter
}

\DeclareAcronym{ecu}{
  short = ECU,
  long  = Electronic Control Unit
}

\DeclareAcronym{nic}{
  short = NIC,
  long  = network interface card
}

\DeclareAcronym{udp}{
  short = UDP,
  long  = user datagram protocol
}

\DeclareAcronym{rtt}{
  short = RTT,
  long  = round-trip time
}

\DeclareAcronym{xdp}{
  short = XDP,
  long  = express data path
}

\DeclareAcronym{ptp}{
  short = PTP,
  long  = Precision Time Protocol
}

\DeclareAcronym{afxdp}{
  short = AF\_XDP,
  long = address family \ac{xdp}
}

\DeclareAcronym{cpu}{
  short = CPU,
  long  = Central Processing Unit
}

\DeclareAcronym{umem}{
  long = Universal Memory Module,
  short = UMEM
}

\DeclareAcronym{dpdk}{
  short = DPDK,
  long  = data plane development kit
}

\DeclareAcronym{tsn}{
  short = TSN,
  long  = time-sensitive networking
}

\DeclareAcronym{qos}{
  short = QoS,
  long  = quality of service
}

\DeclareAcronym{vlan}{
  short = VLAN,
  long  = virtual local area network
}

\DeclareAcronym{pcp}{
  short = PCP,
  long  = priority code point
}

\DeclareAcronym{skb}{
  short = SKB,
  long  = socket buffer
}

\DeclareAcronym{owd}{
  short = OWD,
  long  = one-way delay
}

\begin{document}

\def\enablevspacing{}

\renewcommand{\arraystretch}{1.2}

\newcommand{\EvalDelay}{10ms}
\newcommand{\avgBootTimeRPI}{17.2s}
\newcommand{\avgBootTimeRPIScheme}{21.3s}
\newcommand{\avgBootTimeUltra}{15.3s}
\newcommand{\avgBootTimeUltraScheme}{18.4s}
\newcommand{\avgRelSchemeInc}{23.84}
\newcommand{\avgAbsSchemeIncRPI}{4.1s}
\newcommand{\avgTimeValidation}{306ms}
\newcommand{\avgAbsSchemeIncUltra}{3.1s}
\newcommand{\numSysFilesPIs}{130,578}
\newcommand{\numSysFilesUltrascale}{48,738}
\newcommand{\avgNumBin}{4.3}
\newcommand{\WaitingOverhead}{600ms}
\newcommand{\WaitingOverheadFlat}{1.8s}
\newcommand{\TotalSchemeTime}{21.9s}
\newcommand{\TotalSchemeTimeBus}{23.1s}
\newcommand{\DelayTransmitValidate}{1036ms}
\newcommand{\SizeIntegrityIdentifer}{1130 KB}
\newcommand{\TotalOverhead}{5.236s}
\newcommand{\numECUs}{19}

\newcommand{\SizeCert}{1048 KB}
\newcommand{\SizePK}{1024 KB}
\newcommand{\SizeActionString}{64 B}
\newcommand{\SizeRegistrationRequest}{2136 KB}

\newcommand{\ZoneController}[1]{\ensuremath{\text{ZC}_{#1}}\xspace}

\newcommand{\SoftwareComponents}[1]{\ensuremath{\textsf{SC}_{\footnotesize#1}}}

\newcommand{\AnyECU}[1]{\ensuremath{\ifblank{#1}{\mathcal{E}_\text{i}}{\mathcal{E}_\text{#1}}}\xspace}
\newcommand{\CentralECU}{\AnyECU{C}}
\newcommand{\Adv}{\ensuremath{\mathcal{A}\textit{dv}}\xspace}
\newcommand{\VehicleID}{\ensuremath{\textit{VID}}\xspace}

\newcommand{\GenericToken}[1]{\ensuremath{\textsf{#1}}\xspace}
\newcommand{\Cert}{\ensuremath{\textsf{crt}}\xspace}
\newcommand{\IntID}[1]{\ensuremath{\textsf{IntID}_{\footnotesize #1}}\xspace}
\newcommand{\IntIDChildren}[1]{\ensuremath{\IntID{#1}^{\footnotesize C}}\xspace}
\newcommand{\IM}[1]{\ensuremath{\textsf{IntM}_{\footnotesize #1}\xspace}}
\newcommand{\IMChildren}[1]{\ensuremath{\IM{#1}^{\footnotesize C}}\xspace}
\newcommand{\AttestationKey}{\ensuremath{\mathcal{A}\mathcal{K}\xspace}}
\newcommand{\RootKey}[1]{\ensuremath{\textsf{SK}_{\footnotesize#1}\xspace}}
\newcommand{\Cipher}[2]{\ensuremath{\textsf{C}^{\footnotesize#1}_{\footnotesize#2}}\xspace}
\newcommand{\Nonce}{\ensuremath{\mathcal{N}}\xspace}
\newcommand{\mac}{\ensuremath{mac}\xspace}

\newcommand{\GenericFunc}[2]{\ensuremath{\operatorname{\textit{#1}}\ifblank{#2}{}{(#2)}\xspace}}
\newcommand{\Certify}[1]{\GenericFunc{Cert}{#1}}
\newcommand{\Hash}[1]{\GenericFunc{Hash}{#1}}
\newcommand{\Load}[1]{\GenericFunc{Load}{#1}}
\newcommand{\MAC}[2]{\GenericFunc{MAC}{#1}_{#2}}
\newcommand{\Encrypt}[2]{\GenericFunc{Enc}{#1}_{#2}}
\newcommand{\Decrypt}[2]{\GenericFunc{Dec}{#1}_{#2}}
\newcommand{\SendToParent}[1]{\GenericFunc{SendToParent}{#1}}
\newcommand{\Parent}[1]{\GenericFunc{Prnt}{#1}}
\newcommand{\Measure}{\GenericFunc{Measure}{}}
\newcommand{\Notify}{\GenericFunc{Notify}{}}
\newcommand{\BootCodes}{\GenericFunc{GetAuthBootCodes()}{}}

\newcommand{\Sign}[2]{\ensuremath{\operatorname{Sign}({\footnotesize #1})_{#2}}\xspace}
\newcommand{\Verify}[2]{\ensuremath{\operatorname{Verify}(#1)_{#2}\xspace}}


\newcommand{\Challenge}{c\xspace}
\newcommand{\Signed}[1]{\ensuremath{#1_{\text{sig}}}\xspace}

\newcommand{\hash}{\ensuremath{\texttt{hash}}\xspace}
\newcommand{\authhash}{\ensuremath{\texttt{auth}}\xspace}

\newcommand{\software}{\ensuremath{\mathsf{sw}}\xspace}

\newcommand{\Pk}[1]{\ensuremath{\textsf{pk}_{#1}}\xspace}
\newcommand{\Sk}[1]{\ensuremath{\textsf{sk}_{#1}}\xspace}



\newcommand{\qmarks}[1]{``#1''}
\newcommand{\para}[1]{\smallskip\noindent\textbf{#1.}}

\newcommand{\vspacing}[1]{\ifdefined\enablevspacing\vspace{#1}\fi}

\title{Towards Timing Isolation for Mixed-Criticality Communication in Software-Defined Vehicles}

\author{
	\IEEEauthorblockN{Lóránt Meszlényi\IEEEauthorrefmark{1}$^1$, Julius Kahle\IEEEauthorrefmark{1}$^2$, Dominik P{\"u}llen\IEEEauthorrefmark{1}$^1$, Stefan Kowalewski$^2$,\\Stefan Katzenbeisser$^1$, Alexandru Kampmann\IEEEauthorrefmark{1}$^2$}
	\IEEEauthorblockA{\IEEEauthorrefmark{1}These authors contributed equally to this work.}
	\IEEEauthorblockA{$^1$University of Passau, Germany, Email: \{firstname.lastname\}@uni-passau.de}
	\IEEEauthorblockA{$^2$RWTH Aachen University, Germany, Email: \{lastname\}@embedded.rwth-aachen.de}
}
\maketitle


\fancypagestyle{firstpage}{
  \fancyhf{}
  \renewcommand{\headrulewidth}{0pt}
  \renewcommand{\footrulewidth}{0pt}
  \fancyfoot[C]{\copyright~2025 IEEE. Personal use of this material is permitted. To appear in Proceedings of IAVVC 2025.}
}

\thispagestyle{firstpage}

\begin{abstract}
As the automotive industry transitions toward centralized Linux-based architectures, ensuring the predictable execution of mixed-criticality applications becomes essential. 
However, concurrent use of the Linux network stack introduces interference, resulting in unpredictable latency and jitter.
To address this challenge, we present a layered software architecture that enforces timing isolation for Ethernet-based data exchange between mixed-criticality applications on Linux-based automotive control units. 
Our approach integrates traffic prioritization strategies at the middleware layer, the network stack layer, and the hardware layer to achieve isolation across the full software stack. 
At the middleware layer, we implement a fixed-priority, non-preemptive scheduler to manage publishers of varying criticality. 
At the network layer, we leverage the \ac{XDP} to route high-priority data directly from the network interface driver into critical application memory, bypassing the standard Linux network stack.
At the hardware layer, we dedicate a \ac{NIC} queue exclusively to real-time traffic. 
We demonstrate how our architecture performs in a \Ac{DDS}-based system.
Our evaluation shows that the approach leads to consistent and predictable latencies for real-time traffic, even under heavy interference from best-effort applications.
\end{abstract}

\section{Introduction}
\label{sec:introduction}
Modern automotive software architectures aim for a high degree of flexibility to enable over-the-air updates and to accelerate time-to-market of new features. 
Achieving this flexibility requires appropriate middleware technology, both for task execution and for data communication.
Examples of data-communication middleware that follow the publish–subscribe paradigm include \Ac{DDS} \cite{omgAboutData}, SOME/IP \cite{someip}, and Zenoh \cite{corsaro2023zenoh}.  
They allow distributed applications to exchange data in a flexible manner, since data flows are dynamically established at runtime.
The resulting system consists of software components that form cause-effect-chains, where data propagate through alternating computation and communication phases, forming a compute–communicate–compute graph.

Traditionally, safety-critical and best-effort applications were strictly separated on different execution platforms.
However, the automotive industry is shifting towards centralized \acp{ECU}, which now become systems that execute workloads of mixed criticality \cite{10.1145/3131347}.
Meeting varying levels of timing and safety requirements while sharing resources requires a careful design of the hardware and software architecture.
For real-time systems, both task execution as well as timely data communication between elements of an effect-chain are critical.
Once a critical calculation completes, its results must be reliably transmitted to actuators or other systems within strict timing bounds \cite{Esper}. 

There are ongoing efforts to use Linux as an operating system for safety-critical functions, as a replacement for the proprietary solutions that are traditionally used in these environments.
However, the original Linux kernel architecture prioritized fair resource sharing and overall system throughput rather than the predictable timing behavior required by real-time applications \cite{gemlau2022efficient}. 

\begin{figure}[!t]
	\centering
	\resizebox{\linewidth}{!}{\includegraphics{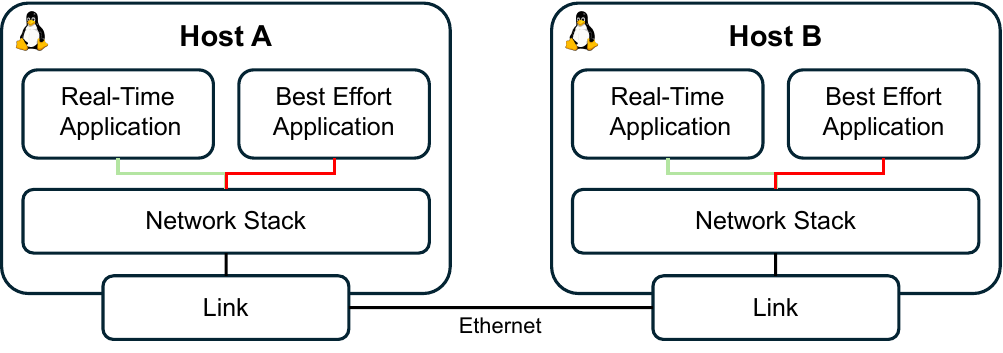}}
	\caption{Our objective is to leverage standard Linux scheduling and networking mechanisms to minimize the impact of best-effort traffic on the jitter and maximum end-to-end latency experienced by the real-time traffic.}
	\label{fig:problem_description}
\end{figure}
Our work focuses on communication characteristics between mixed-criticality workloads that are executed on Linux.
For this purpose, we consider a minimal testbed of two Linux hosts, as depicted in \cref{fig:problem_description}.
We generate interference traffic at line-rate between them to emulate \textbf{best-effort} applications. 
Parallel to that background load, each host runs a \textbf{real-time} application which exchanges messages with its peer. 
Our goal is to minimize the impact of \textbf{best-effort} traffic on the jitter and maximum latencies of the communication between the \textbf{real-time} applications while using existing Linux mechanisms.

We address this challenge through a layered architecture that combines priority-based scheduling of middleware operations, kernel-bypassing techniques and careful network hardware configuration to reduce the effect of concurrent interference traffic.
Our contributions are as follows:
\begin{itemize}
	\item we describe sources of interference at each layer of the stack,
	\item we present a layered architecture that implements strategies to mitigate interferences, and
	\item we demonstrate bounded latencies for real-time traffic despite heavy interference traffic in experiments.
\end{itemize}
Our paper is structured as follows.
Section 2 presents related work.
Section 3 describes various interference along the full software and hardware stack. 
We describe our layered architecture and the mitigation strategies we employed.
Section 4 presents our evaluation results, including latency, jitter and performance under various load conditions.
Section 5 discusses the implications of our findings and potential applications in industrial and automotive domains.
\section{Related Work}
\label{sec:related_work}

We focus particularly on approaches that address timing isolation and deterministic networking.

\subsection{XvSomeIP: A High-Performance In-Vehicle Communication Middleware}
\citeauthor{10774081} propose XvSomeIP, a high-performance implementation of the SOME/IP (Scalable service-Oriented Middleware over IP) protocol leveraging eXpress Data Path (XDP) framework to enhance communication efficiency in automotive systems. Similar to our approach, their work addresses the real-time performance limitations of traditional in-vehicle network communication.

XvSomeIP integrates XDP with the vSomeIP middleware to bypass the kernel networking stack, thereby reducing packet processing latency and increasing throughput. By implementing packet encapsulation, filtering, and decomposition in user space, XvSomeIP avoids the overhead associated with interrupt triggering, memory copying, and system calls that typically impact traditional middleware performance.

The authors demonstrate significant performance improvements, including a \SI{30}{\percent} reduction in packet processing latency and a $4\times$ increase in throughput when handling 1024-byte packets. Their architecture employs a CPU-friendly memory model and user-space packet processing to achieve these gains.

While our work shares similar technologies for optimizing Ethernet communication through kernel bypass techniques, XvSomeIP focuses specifically on the SOME/IP protocol with XDP as its core technology.

In contrast, our approach implements a hierarchical task management system that separates safety-critical and mission-critical tasks from best-effort tasks, using a dedicated userspace network stack for priority tasks while allowing best-effort tasks to utilize the conventional OS network stack.

Further, we apply configuration to \ac{nic} and Linux queueing disciplines in order to configure the network.
While XvSomeIP primarily targets overall communication performance improvement by only implementing custom networking stack, our solution takes a step back and handles the challenge of reduced latency and jitter communication by a holistic solution.

\subsection{INSANE Framework for Ultra-Low Latency Communication}
The INSANE (Intelligent Networking for Synchronous Applications) Framework \cite{insane} is a communication middleware designed to achieve ultra-low latency in datacenter applications. It operates in userspace and employs lockless queues, offering different adapters for communication such as \ac{afxdp}, \ac{dpdk}, and \Ac{rdma}. The framework includes its own packet scheduler that manages packets based on their quality of service (QoS) properties.

While the INSANE framework demonstrates impressive performance metrics, several limitations must be considered in the context of automotive architectures:

\subsubsection{Hardware Dependency}
The evaluation of INSANE relies heavily on high-performance components like the Mellanox DX-5/6 \SI{100}{Gbits} \acp{nic}. 
These \acp{nic} are equipped with offloading techniques that INSANE depends on. This hardware dependency limits the applicability of INSANE to environments where such specialized \acp{nic} are not available, making it less suitable for standard automotive \acp{ecu}.

\subsubsection{Packet Scheduling}
Since INSANE is designed to be protocol-agnostic, its packet scheduling logic is positioned later in the processing pipeline. This design restricts the flexibility of influencing scheduling decisions from user software and requires all packets to be handled by the packet scheduler, regardless of their priority or QoS properties. In a mixed-criticality \ac{ecu} context, this limitation can be problematic as it does not allow for dynamic prioritization based on real-time requirements.

\subsubsection{Dependency on DPDK for Optimal Performance}
INSANE achieves optimal performance with \ac{dpdk}, a userspace networking framework tailored for high-performance networking. When using \ac{dpdk}, the \ac{nic} is removed from the control of the operating system, and the userspace program occupies the entire \ac{nic}. This can be a significant drawback, since all required network services must be reimplemented in userspace. This limitation also implies that all processes on the \ac{ecu} that need communication are required to be managed by INSANE.

Moreover, the experimental validation of INSANE does not account for protocol processing overhead. 
In contrast to this, we implement the experimental verification on a real-world usecase.
We use the Linux \ac{afxdp} framework for bypassing the kernel, which offers better flexibility in handling mixed-traffic scenarios compared to \ac{dpdk}. 
This approach aligns more closely with the requirements of automotive architectures, where both best-effort and high-priority traffic must coexist efficiently.

\subsection{Real-Time Networking for DDS Middleware}
\subsubsection{Advancing User-Space Networking for DDS}
Another relevant study by \citeauthor{10494460} \cite{10494460} focuses on extending the \ac{dds} for real-time capabilities. The authors address the issue of latency-critical applications over the \ac{dds} standard by enhancing CycloneDDS \cite{cycloneddsEclipseCyclone} with two main userspace networking techniques: \ac{afxdp} and \ac{dpdk}. Their implementation achieves a 94\% reduction in the mean latency bound and an 18\% reduction in overall latency by using \ac{afxdp} compared to the standard Linux network stack.

\subsubsection{Real-Time Capability of DDS Implementations}
In an additional investigation, \citeauthor{BODE2025102013} \cite{BODE2025102013} examine the real-time capability of four different \ac{dds} implementations in combination with userspace networking libraries. 
They explicitly exclude the usage of \ac{tsn}, as this requires extensive hardware support for deterministic Ethernet behavior. 
The study evaluates the performance of CycloneDDS, OpenSplice, and RTI Connext under various networking conditions. 
The results show that while userspace networking techniques can improve latency, achieving hard real-time guarantees remains challenging without specialized hardware support.

Unlike these approaches that focus on enhancing existing middleware solutions, our work proposes a comprehensive architecture that addresses both execution within the middleware and network communication, providing stronger isolation guarantees for mixed-criticality systems.
Different to the evaluation methodology of \citeauthor{BODE2025102013}, we measure the effectiveness of our proposed solution also with the presence of interference traffic on the same link, mimicking real world use-case scenarios. 

\subsection{Efficient Timing Isolation for Mixed-Criticality Communication Stacks in Performance Architectures}
The mixed-criticality communication architecture proposed by \citeauthor{9921445} differs fundamentally from our \ac{afxdp} implementations in several aspects. 
While both approaches aim to improve networking performance for time-critical applications, their underlying mechanisms vary significantly.
\citeauthor{9921445} utilize physical separation through heterogeneous computing architecture, dedicating hardware resources to critical traffic processing.
In contrast, an \ac{afxdp} approach maintain processing within the Linux environment, employing kernel-bypass techniques while sharing underlying hardware resources. 
Although \ac{afxdp} reduces software overhead through zero-copy operations and direct NIC access, it lacks the physical isolation guarantees offered by \citeauthor{9921445}'s solution.
In contrast to this contribution, our solution relies on the isolation mechanism provided by the \ac{nic} instead of custom hardware platform.

\section{Timing Isolation Architecture}
\label{sec:integrity_validation_scheme}
\begin{figure}[!t]
    \centering
    \resizebox{\linewidth}{!}{\includegraphics{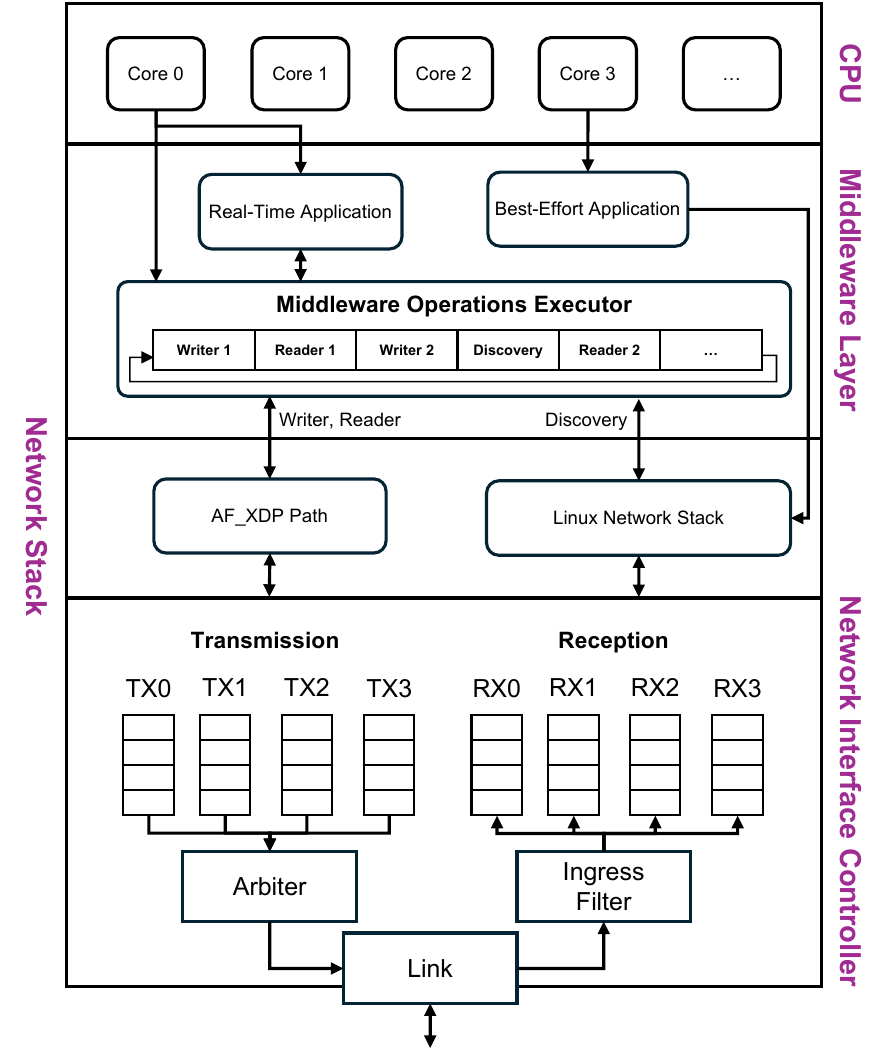}}
    \caption{We propose mitigation at different technical levels to reduce the impact of interference traffic on concurrent real-time traffic. We address the middleware layer, the network stack, as well as the network interface controller.}
    \label{fig:architecture}
\end{figure}

We will now describe various sources of interference across the full software and hardware stack and the strategies we implemented to mitigate them. 
As depicted in \cref{fig:architecture}, we incorporate the middleware layer, the network stack, and the hardware layer.
Additionally, we isolate real-time and best-effort workloads on individual isolated CPU cores.

\subsection{Middleware Layer}
\textbf{Challenge.} In addition to the transmission and reception of user traffic, the middleware layer handles a manifold of tasks, such as discovery, liveliness monitoring of remote peers or the processing of retransmission requests.
These operations are of varying criticality.
For example, the handling of housekeeping operations is less important than delivery of critical user traffic.
Additionally, there might be multiple publishers and subscribers with different priority levels.
For example, in \Ac{DDS}, there are different flows for user traffic and discovery traffic (called \textit{meta traffic} in DDS terminology).

At this layer, our objective is to guarantee a predictable and deterministic execution of the operations executed at the middleware layer - regardless of the other concurrent operations that have to be executed at this layer.
Specifically, we aim to ensure that the transmission of user data has a known and fixed response time, irrespective of other pending operations.
Similarly, the processing of incoming user traffic shall not be deferred in favor of lower priority traffic.
We believe that achieving this guarantee is considerably more challenging in multithreaded, lock-based systems.

\textbf{Our approach.} Our prototypical \ac{DDS} implementation is single threaded, and is based on embeddedRTPS  \cite{kampmann2019portable}.
The single thread drives an executor that handles all operations occurring at the middleware layer.
This executor acts as a non-preemptive, fixed-priority scheduler for all middleware operations.
If multiple operations are pending concurrently, we can explicitly prioritize critical operations.
When multiple operations are pending concurrently, we can explicitly prioritize critical ones. This approach allows us to bound the maximum response time of pending operations: a critical operation is only delayed up to the maximum worst-case execution time of any middleware operation.

We make use of the \texttt{epoll} mechanisms to avoid busy polling.
The executor operates event and time-triggered, handling all middleware operations, such as transmission of user traffic, reception, and processing of incoming user and discovery packets, periodic transmission of discovery messages, as well as periodic monitoring of remote peer liveliness.
When no tasks are pending, the thread suspends itself to avoid consuming further computation time.
Additionally, our single-thread implementation eliminates the need for locking mechanisms.
\subsection{Network Stack}
\textbf{Challenge.}
The Linux network stack is designed for throughput rather than predictable timing \cite{gemlau2022efficient}.
For example, it switches between polling and interrupted-driven mode in dependence of network load.
Addition, communication requires system calls for transmission and reception of messages.
These not only introduce significant overhead, they also require copying data between kernel and userspace (e.g., \texttt{recvfrom}, \texttt{sendto}).
Additionally, packets traverse multiple processing layers and queueing disciplines before finally being copied from the system memory to one of the \ac{nic}'s queues using \Ac{dma}.
In doing so, best-effort and real-time traffic traverse common data structures and memory pools that are protected by global spinlocks \cite{10.1145/3452296.3472888}, where contention can lead to mutual interference between traffic classes.

\textbf{Our approach.} 
Our implementation uses the Linux \ac{afxdp} framework to bypass large parts of the network stack.
\ac{afxdp} operates by attaching \ac{ebpf} programs at an early point in the ingress path of the \ac{nic}.
The \ac{afxdp} socket framework leverages a shared memory region, referred to as \Ac{umem}, which is accessible to both the kernel and userspace.
\Ac{umem} is structured as a lock-free ring buffer, comprising four distinct queues: the transmit (TX) queue, the receive (RX) queue, the fill queue (FQ), and the completion queue (CQ).
This avoids overhead associated with packet copying, as both kernel and userspace components directly operate on the same memory region.
By bypassing the traditional kernel networking stack, the \ac{afxdp} framework achieves reductions in latency and jitter, while maintaining compatibility with widely available hardware. 

For each received packet, the loaded \ac{ebpf} program determines how the packet is processed: it is either forwarded to the regular Linux networking stack, dropped, redirected to a different \ac{NIC} or directly passed to a user space application. 
We use the latter mechanism to redirect incoming traffic right into our middleware layer in a zero-copy fashion.

The \ac{ebpf} program only forwards \Ac{udp} packets that contain the RTPS magic numbers to the middleware layer. 
All other packets are forwarded to the regular network stack, making our \ac{afxdp} program transparent to other best-effort applications.
A drawback of using \ac{afxdp} is that we have to manually craft the full Ethernet frame, including UDP/IP headers.
\subsection{Network Interface Controller}
\textbf{Challenge.}
As depicted in \cref{fig:architecture}, modern \ac{nic} have multiple queues for transmission and reception of traffic.
The Intel i226-based \ac{nic} used in our experiments has four separate transmission (TX0-3) and reception (RX0-3) queues.

On the ingress side, the \ac{nic} assigns packets to one of the RX queue and copies it via \ac{DMA} into memory.
An interrupt triggered on one of the CPU cores notifies the kernel of the new packet.
If real-time packets share an RX queue with best-effort traffic, they are serviced only after the driver has drained all preceding frames, significantly increasing their queuing delay.

For transmission, if multiple TX queues have packets ready for transmit, an arbiter decides which queue to service first.
In case of the Intel i226, the arbiter operates in a fixed-priority manner, with TX-0 having the highest priority, and TX-3 having the lowest priority.
As on the ingress side, real-time traffic can be significantly delayed if mapped to a low-priority queue, or if real-time traffic shares a queue with best-effort traffic.

\textbf{Our approach.}
Similar to \cite{gemlau2022efficient}, we configure the \ac{nic} to exclusively reserve queues for priority traffic, ensuring separation from best-effort communication.
More specifically, we always map real-time traffic to RX-0 and TX-0.
Best-effort traffic is distributed among the remaining queues.

On the ingress side, we use the packet filtering mechanism offered by the Intel i226 to achieve this separation. 
Based on the \Ac{vlan}-tag in the IEEE 802.1Q Ethernet header, the \Ac{NIC} maps the packet to 
queue RX-0, to which also our \ac{afxdp} socket is bound.
This mechanism is configurable through \texttt{ntuple} filters in Linux.

To isolate egress traffic, we bind the \ac{afxdp} socket to the highest priority queue (TX-0) of our \ac{nic}.
This is sufficient to steer our \Ac{afxdp} traffic to TX-0.
However, our \ac{afxdp} socket only handles unicast user traffic.
We steer the remaining meta traffic to TX-0 via the \texttt{mqprio} queuing discipline.
This traffic is considered as best-effort and traverses the full network stack.
\texttt{mqprio} queueing discipline allows to steer traffic to specific hardware queues based on the \Ac{skb} priority, which we assign to the meta traffic.
Traffic originating from other processes or kernel threads is not assigned a high \ac{skb} priority.
Thus, it is automatically distributed to TX-1 to TX-3.

The default egress queuing discipline is \texttt{fq\_codel}, which is an algorithm that aims to give fair queueing to different \textit{flows} within the same traffic class.
We replace \texttt{fq\_codel} with \texttt{pfifo}, that implements a simple first-in first-out queueing discipline to reduce complexity.
Note that \ac{afxdp} bypasses both \texttt{fq\_codel} and \texttt{mqprio}.
In a \ac{TSN} setup, a traffic shaper is installed on TX-0 to avoid starvation of lower priority queues \cite{finn2018introduction}.
We omit this step for simplicity in our setup. 

Finally, we also disable interrupt coalescing on the \ac{nic} to avoid delays of RX interrupts.
\section{Evaluation}
\label{sec:evaluation}
\begin{figure}[!t]
	\centering
	\resizebox{0.7\linewidth}{!}{\includegraphics{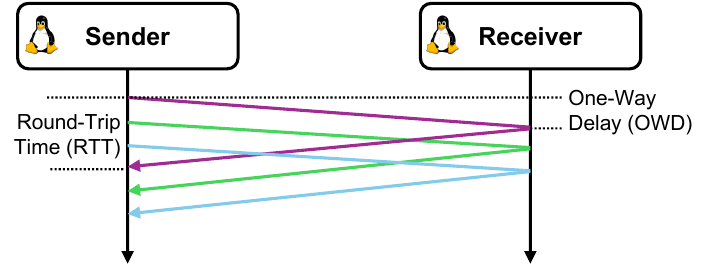}}
	\caption{We measure \acfp{rtt} and \acfp{owd} in an open-loop fashion, not waiting for the response.
		Real-time packets are generated at \SI{2}{\kilo\hertz} with varying packet sizes. }
	\label{fig:latency_measurement}
\end{figure}
\acreset{rtt}\acreset{owd}

We evaluate the presented approach for in a series of experiments.
In \cref{sec:evaluation:setup} we describe the experimental setup, including both the scenarios under test and the hardware and software configurations used in the evaluation.
In \cref{sec:evaluation:results} we present the results of our experiments, comparing jitter and latencies of the presented approach against the baseline configuration.
\cref{tab:udp} and \cref{tab:xdp} show the results for all our experiments.
In \cref{sec:evaluation:discussion} we discuss the our results.
\subsection{Setup}
\label{sec:evaluation:setup}

Our testbed consists of  two Dell Precision 3680 computers equipped with an Intel i9-14900 CPU and \SI{64}{\giga\byte} of RAM.
The computers are connected over a direct link using Intel I226 \acp{nic}.
We synchronize time between both machines using \Ac{PTP} to allow for comparison between timestamps captured on different machines.
Both machines run Ubuntu 24.04.2, kernel version 6.8.1-1019-RT with PREEMPT-RT patch.

One machine acts as the sender and the other as the receiver, as depicted in \cref{fig:latency_measurement}.
Both sender and receiver run instances of our prototypical \ac{dds} implementation.
The sender uses a dedicated thread to publish messages on the topic \texttt{RTTPing} at every \SI{500}{\micro\second}, and a subscriber that receives messages on the topic \texttt{RTTPong}.
The receiver has a publisher that sends messages on the topic \texttt{RTTPong} and a subscriber that receives messages on the topic \texttt{RTTPing}.
We send messages in an open-loop fashion and do not wait for the response from the other machine.

A callback on the receiver side handles the incoming message and immediately sends a reply on the topic \texttt{RTTPong}.
This message is received by a callback on the sender side, which completes the round-trip.
We use LTTng as a tracing framework to capture timestamps.
Additionally, the packets payload size is varied between \SI{64}{\byte}, \SI{512}{\byte}, and \SI{1024}{\byte}.

We have conducted experiments for two configurations.
At the middleware layer, the setup is identical for all experiments and uses our single-thread \Ac{DDS} implementation.
For the \textbf{baseline configuration}, we do not employ any of the mitigation strategies at the network stack and at the \ac{NIC} level.
Instead of \ac{afxdp}, we utilize conventional socket programming for data transmission, and packets traverse the default networking stack with default configuration.

The second configuration employs all proposed mitigation strategies, making use of \Ac{afxdp} as well as the queue separation at the \ac{nic} level.

In both configurations, the sender and receiver processes running on each machine are pinned to core 1 and are scheduled using \texttt{SCHED\_FIFO} policy at priority 99.

We study the influence of interference traffic on \ac{rtt} and \ac{owd} for both configurations.
Interference traffic is generated using the traffic generator \texttt{iperf3}.
The \texttt{iperf3} server runs on the receiver, to which the \texttt{iperf3} client executed on the sender machine connects.
We run \texttt{iperf3} in \textbf{bidirectional} mode, with $50$ parallel connections with a target bandwidth of \SI{2.5}{Gbits}, which is the line rate of the I226 \Ac{NIC}.
\texttt{iperf3} is configured to generate UDP traffic, as running \texttt{iperf3} in the default TCP mode triggers congestion control, which unintentionally reduces the amount of interference traffic.
The \texttt{iperf3} payload size is identical to the payload size of our \ac{DDS} traffic.
We confirmed that \texttt{iperf3} reaches the available line rate, when considering the traffic generated by our \Ac{dds} application.
\subsection{Results}
\label{sec:evaluation:results}
\begin{table*}
  \centering
  \caption{Results using the baseline configuration (all times in \unit{\micro\second}).}
  \begin{tabular}{llllllllll}
\toprule
Payload & RTT max. & OWD max. & RTT min. & OWD min. & RTT var. & OWD var. & RTT med. & OWD med. & Interf. \\
\midrule
1024B & $1141.1$ & $335.67$ & $190.32$ & $96.64$ & $895.68$ & $9.86$ & $349.78$ & $250.55$ & off \\
512B & $1214.82$ & $328.43$ & $165.8$ & $133.92$ & $88.1$ & $8.66$ & $386.88$ & $247.81$ & off \\
64B & $560.46$ & $264.58$ & $74.37$ & $46.01$ & $1633.95$ & $10.68$ & $363.07$ & $244.3$ & off \\
\midrule
1024B & $3285.33$ & $2498.36$ & $1796.11$ & $1393.67$ & $13055.28$ & $4374.07$ & $2161.61$ & $1649.88$ & on \\
512B & $3340.67$ & $1837.19$ & $1771.91$ & $1340.05$ & $148494.1$ & $4770.94$ & $2160.52$ & $1624.59$ & on \\
64B & $3149.94$ & $1807.29$ & $685.91$ & $335.04$ & $115867.49$ & $46235.09$ & $2041.96$ & $1558.13$ & on \\
\bottomrule
\end{tabular}

  \label{tab:udp}
\end{table*}

\begin{table*}
  \centering
  \caption{Results using our timing isolation architecture (all times in \unit{\micro\second}).}
  \begin{tabular}{llllllllll}
\toprule
Payload & RTT max. & OWD max. & RTT min. & OWD min. & RTT var. & OWD var. & RTT med. & OWD med. & Interf. \\
\midrule
1024B & $448.87$ & $300.32$ & $182.9$ & $139.74$ & $11.02$ & $7.34$ & $397.89$ & $251.36$ & off \\
512B & $431.65$ & $288.81$ & $172.05$ & $136.32$ & $9.67$ & $4.62$ & $391.55$ & $248.19$ & off \\
64B & $480.86$ & $270.13$ & $235.0$ & $140.55$ & $6.73$ & $3.42$ & $385.54$ & $245.19$ & off \\
\midrule
1024B & $195.63$ & $114.81$ & $144.19$ & $70.96$ & $41.08$ & $23.39$ & $158.64$ & $81.82$ & on \\
512B & $197.01$ & $118.52$ & $140.23$ & $68.8$ & $50.03$ & $24.85$ & $155.91$ & $80.18$ & on \\
64B & $196.27$ & $114.59$ & $136.27$ & $61.74$ & $36.61$ & $29.98$ & $156.48$ & $77.12$ & on \\
\bottomrule
\end{tabular}

  \label{tab:xdp}
\end{table*}

\begin{figure*}
    \centering
    \begin{subfigure}{.49\linewidth}
      \centering
	  \includegraphics[width=\linewidth]{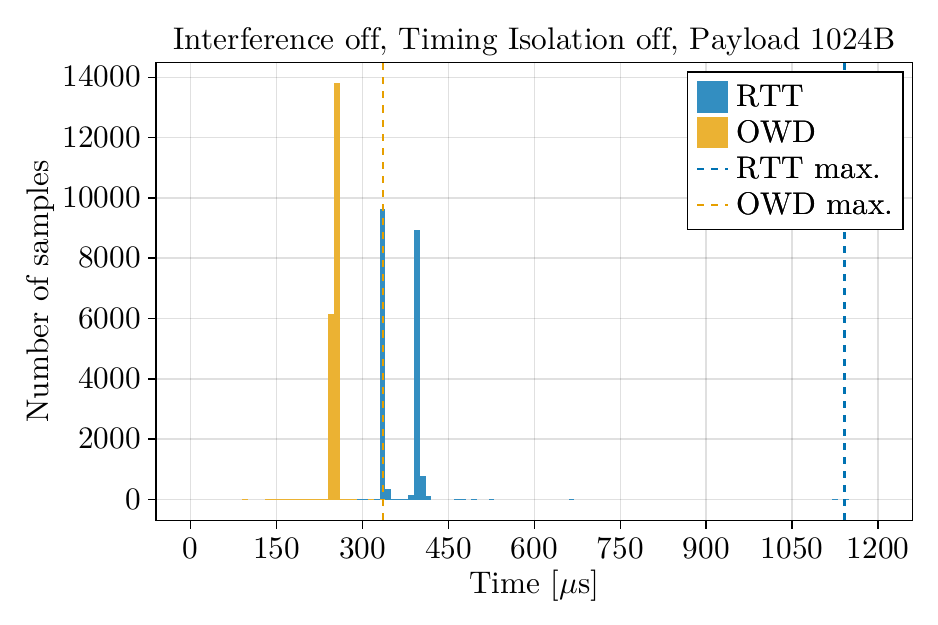}
      \caption{Without interfering traffic.}
      \label{fig:udp_iperf:off}
    \end{subfigure}
    \begin{subfigure}{.49\linewidth}
      \centering
      \includegraphics[width=\linewidth]{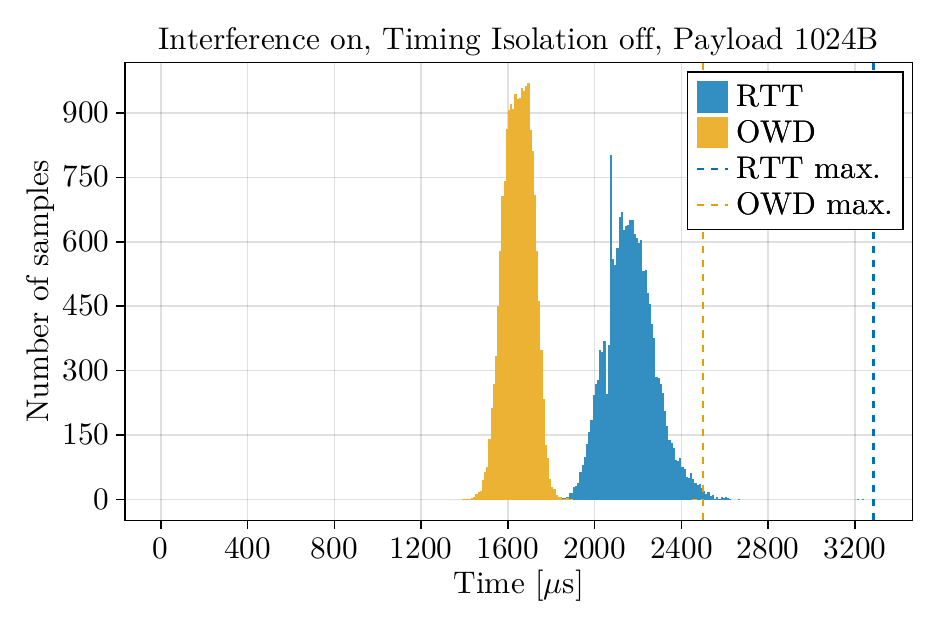}
      \caption{With interfering traffic.}
      \label{fig:udp_iperf:on}
    \end{subfigure}%
    \caption{Distribution and maximum of \acfp{owd} and \acfp{rtt} using the baseline configuration with a payload size of \SI{1024}{\byte}; bin size equals \SI{10}{\micro\second}.}
    \label{fig:udp_iperf}
\end{figure*}

\Cref{fig:udp_iperf} shows the \acp{owd} and \acp{rtt} using the \textbf{baseline} configuration with and without interfering traffic.
\Cref{fig:udp_iperf:off} shows the latency distribution without interference traffic. 
The measured \acp{owd} are tightly distributed, while the \acp{rtt} show slightly more variance.
However, as depicted in \Cref{fig:udp_iperf:on}, we observe a strong performance degradation in the presence of interference traffic.
The variance increases significantly, and the maximum \acp{rtt} increase up to factor $7$.
\begin{figure*}
    \centering
    \begin{subfigure}{.49\linewidth}
      \centering
      \includegraphics[width=\linewidth]{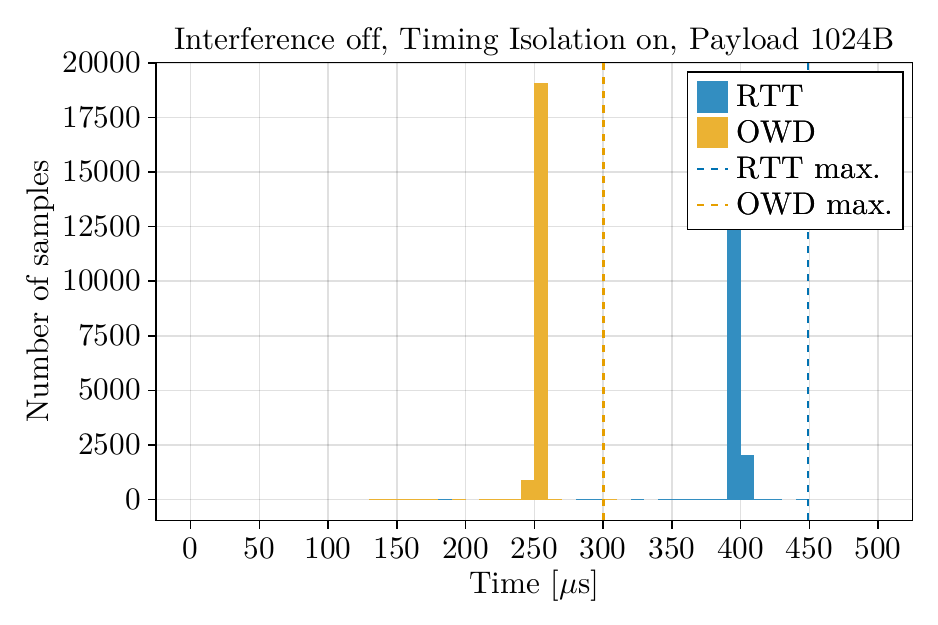}
      \caption{Without interfering traffic.}
      \label{fig:xdp_iperf:off}
    \end{subfigure}
    \begin{subfigure}{.49\linewidth}
      \centering
      \includegraphics[width=\linewidth]{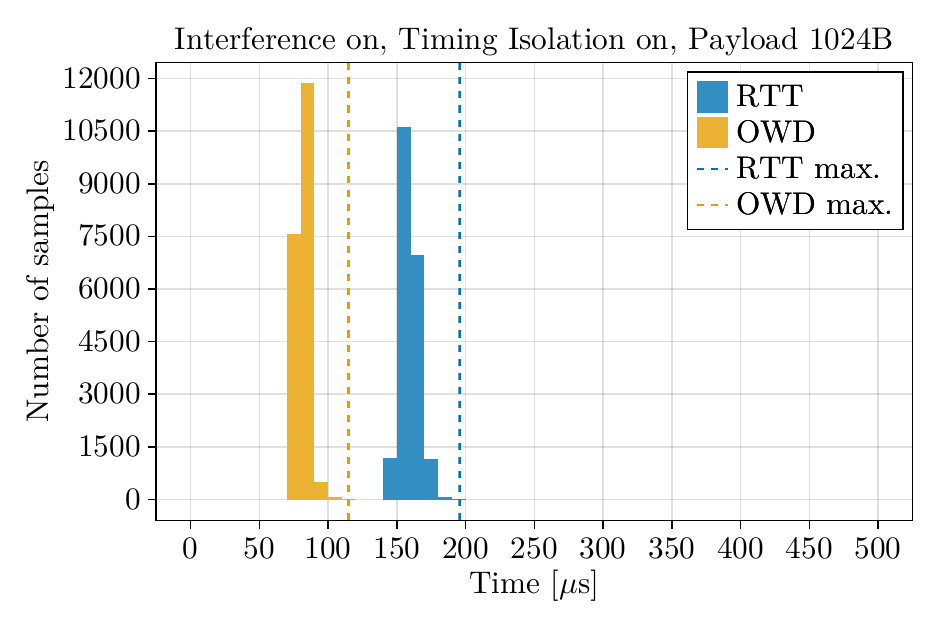}
      \caption{With interfering traffic.}
      \label{fig:xdp_iperf:on}
    \end{subfigure}
    \caption{Distribution and maximum of \acfp{owd} and \acfp{rtt} using our timing isolation architecture with a payload size of \SI{1024}{\byte}; bin size equals \SI{10}{\micro\second}.}
    \label{fig:xdp_iperf}
\end{figure*}

\begin{figure*} 
  \centering
  \begin{subfigure}{.49\linewidth}
    \centering
    \includegraphics[width=\linewidth]{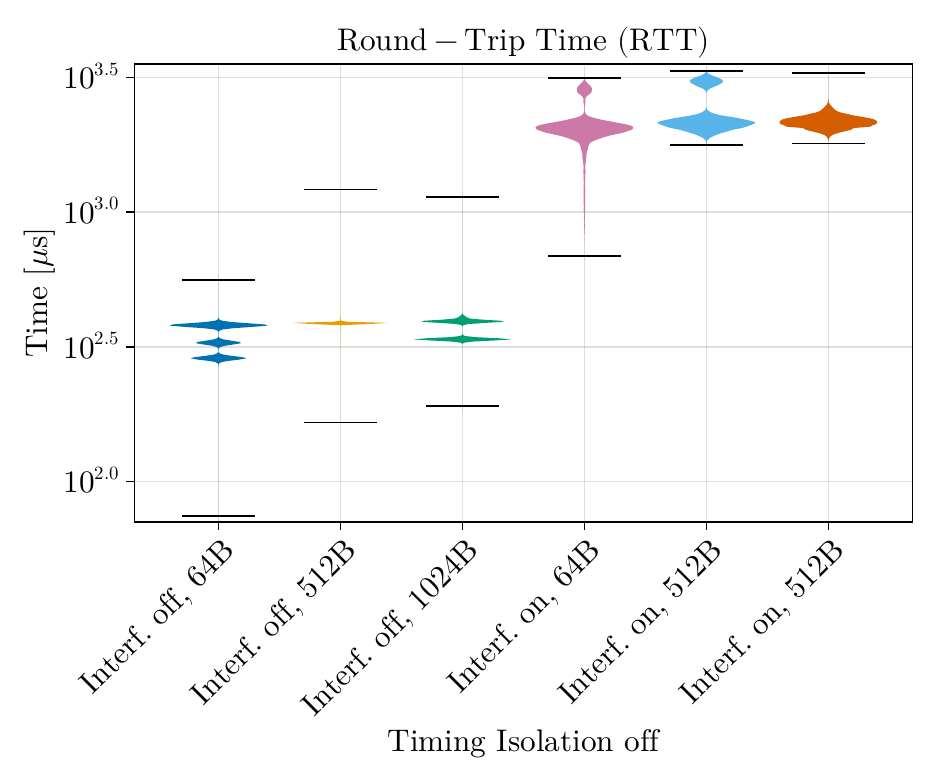}
    \caption{Baseline configuration.}
    \label{fig:rtt:udp}
  \end{subfigure}
  \begin{subfigure}{.49\linewidth}
    \centering
    \includegraphics[width=\linewidth]{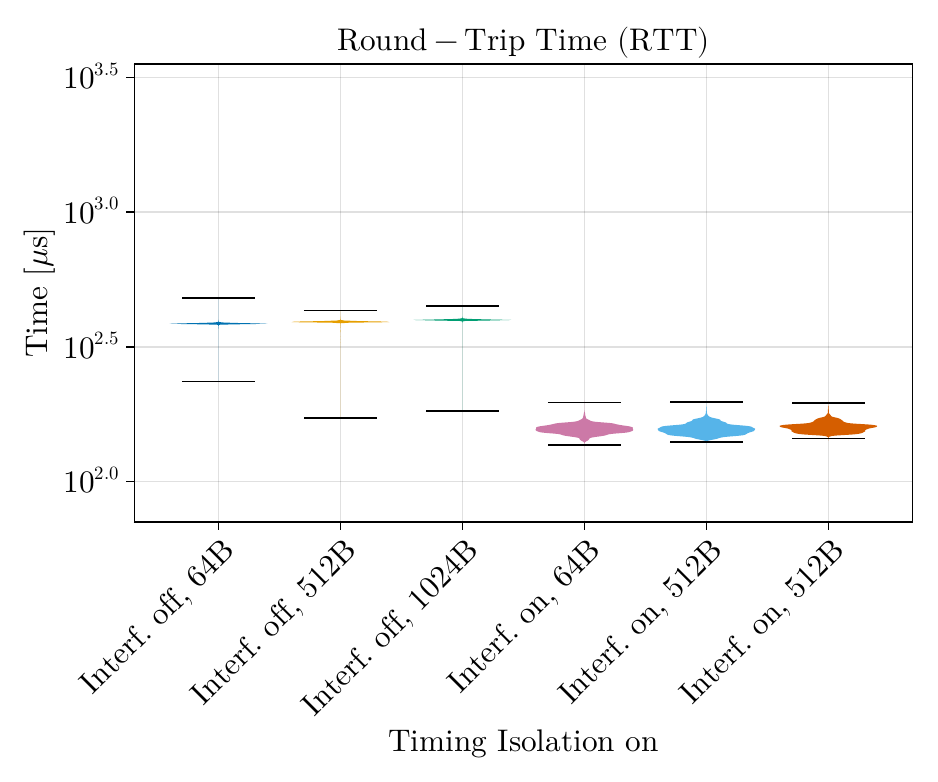}
    \caption{Our approach.}
    \label{fig:rtt:xdp}
  \end{subfigure}
  \caption{Logarithmic distribution and limit values of \acfp{rtt} using the baseline configuration and our timing isolation architecture with different payload sizes, with and without interfering traffic.}
  \label{fig:rtt}
\end{figure*}

In comparison, \cref{fig:xdp_iperf} shows results for \textbf{our approach} with all mitigation strategies enabled.
\cref{fig:xdp_iperf:off} shows the results without interference.
The maximum latencies are not significantly lower than in the baseline configuration.
The latencies become more consistent with significantly lower variance and smaller maximum values.

\Cref{fig:xdp_iperf:on} depicts the results in the presence of interference traffic.
Although we still see an increase in variance, the variance increase is still significantly smaller compared to the baseline configuration. 
Regarding the absolute numbers, we were surprised to see that the latencies \textit{decrease} \cref{fig:xdp_iperf:on} further in the presence of interfering traffic, becoming even lower than the latencies in \cref{fig:xdp_iperf:off} (baseline, no interference).
We suspect that the interference traffic keeps the driver or network stack in a polling mode and prevents it from falling back to an interrupt driven mode.

\Cref{fig:rtt} presents an overview of our evaluation results.
\Cref{fig:rtt:udp} shows the \acp{rtt} using the baseline configuration while \Cref{fig:rtt:xdp} shows the \acp{rtt} using our approach with and without interfering traffic.
We can derive from \cref{fig:rtt} that for every experiment the latency variance decreases when using our approach compared to the baseline configuration.
Further, the mean latencies increases with increasing payload size, which is expectable.
While the latencies increase under the presence of interfering traffic in the baseline configuration, the latencies using our approach decrease under interference of best-effort traffic.

\Cref{tab:udp} and \cref{tab:xdp} shows the key results of our experiments.
The tables show the minimum, median and maximum \acp{rtt} and \acp{owd} and their variance for the baseline configuration and our approach with and without interfering traffic, respectively.
\subsection{Discussion}
\label{sec:evaluation:discussion}

Our experimental results demonstrate the effectiveness of our timing isolation architecture. 
In contrast to related work, we have benchmarked our proposed architecture using line rate interference traffic.
\Cref{tab:udp} shows that the baseline configuration only outperforms our approach in a few cases with respect to minimum \acp{rtt} and. \acp{owd}.
For example, in the case of \SI{64}{\byte} packets without interfering traffic, the minimum \ac{rtt} is \SI{74.37}{\micro\second} compared to \SI{235}{\micro\second} in our approach.
However, the maximum \ac{rtt} -- which is a key performance metric in real-time applications -- is higher for the baseline configuration, reaching \SI{560.46}{\micro\second} compared to \SI{480.46}{\micro\second} in our approach.
Overall, the median values in the experiment without interference traffic are comparable between both configurations.

Our approach not only improves the maximum observed latencies but also the variance, thus leading to a more predictable system behavior.
While the baseline configuration shows a variance of \acp{rtt} of \SI{148494.1}{\micro\second} for \SI{512}{\byte} packets under interfering traffic, our approach achieves a variance of \SI{50.03}{\micro\second} in the same experiment.

We plan to evaluate the effect of the individual isolation strategies in future work.

\section{Conclusion}
\label{sec:conclusion}
We have presented a timing isolation architecture for \ac{xdp}-based communication that significantly improves latency characteristics in real-time applications.
We propose a layered architecture that combines middleware, network stack, and hardware isolation techniques to achieve predictable and low-latency communication.
Our architecture isolates real-time at hardware level, employs \ac{afxdp} to bypass the kernel's networking stack and prioritizes middleware operations with a dedicated executor approach.
Our experimental evaluation demonstrates that the proposed architecture maintains low-latencies even under heavy interference traffic.
The predictable and low-latency performance of our timing isolation architecture offers several benefits for real-time systems:
\begin{itemize}
\item \textbf{Improved determinism}: The significant reduction in latency variance leads to more predictable system behavior, which is crucial for safety-critical applications.
\item \textbf{Robustness under load}: Our architecture maintains consistent performance even under high network utilization, with maximum RTT values remaining under \SI{500}{\micro\second} during network congestion tests.
\item \textbf{Practical deployment}: Our approach achieves performance comparable to specialized solutions while maintaining compatibility with standard kernel infrastructure.
\end{itemize}
In conclusion, our timing isolation architecture provides a practical approach for achieving predictable, low-latency communication using XDP technology, demonstrating performance characteristics suitable for demanding real-time applications.

\section*{Acknowledgement}
\small{
This work has been accomplished within the projects \qmarks{autotech.agil} (FKZ\-01IS22088X) and \qmarks{6G-RIC} (16KISK034). We acknowledge the financial support for the projects by the Federal Ministry of Research, Technology and Space (BMFTR), former Federal Ministry of Education and Research (BMBF).
}

\bibliographystyle{IEEEtranN}
\bibliography{literature}

\end{document}